\documentclass{article}
\usepackage{buckow}
\usepackage{amsmath,amssymb,eucal,bbold}
\usepackage{epic,eepic}
\renewcommand{\d}{\partial}
\newcommand\la{\langle} 
\newcommand\ra{\rangle}
\newcommand{\half}{\tfrac12}
\newcommand{\fg}{\mathfrak g}
\newcommand{\fh}{\mathfrak h}
\newcommand{\ft}{\mathfrak t}
\newcommand{\id}{\mathbb 1}
\newcommand{\Z}{\mathbb Z}
\newcommand{\su}{\mathfrak{su}}
\newcommand{\eC}{\mathcal{C}}
\newcommand{\eM}{\mathcal{M}}
\newcommand{\eZ}{\mathcal{Z}}
\newcommand{\G}{\mathbf{G}}
\newcommand{\U}{\mathrm{U}}
\newcommand{\SU}{\mathrm{SU}}
\newcommand{\SO}{\mathrm{SO}}
\DeclareMathOperator{\Ad}{Ad}

\renewcommand{\Im}{\text{Im}}
\begin{document}
\def\titleline{ A geometric approach to D-branes in group manifolds
  \footnote{SPIN-2001/20, Imperial/TP/01-2/10, hep-th/0112130}}
\def\authors{Sonia Stanciu\1ad\2ad} \def\addresses{\1ad Spinoza
  Institute, Utrecht University, Leuvenlaan
  4, 3508 TD Utrecht,\\ The Netherlands\\
  \2ad Theoretical Physics Group, Blackett Laboratory, Imperial
  College,\\ Prince Consort Road, London SW7 2BZ, U.K.\\
  \texttt{s.stanciu@phys.uu.nl, s.stanciu@ic.ac.uk} }
\def\abstracttext{
  This is a brief review\footnote{Based on talks given at the
    conference \emph{Modern trends in string theory} (Lisbon, July
    2001), and at the RTN network meeting \emph{The quantum structure
      of spacetime and the geometric nature of fundamental
      interactions} (Corfu, September 2001).} of some recent results on
  the geometric approach to symmetric D-branes in group manifolds,
  both twisted and untwisted.  We describe the geometry of the gluing
  conditions and the quantisation condition in the boundary WZW model,
  and we illustrate this by determining the consistent twisted and
  untwisted D-branes in the Lie group $\SU_3$.  } \large \makefront

\section{Introduction}

D-branes in group manifolds have attracted a great deal of attention
in recent years, as they provide an ideal laboratory for the study 
of D-branes in general string backgrounds.  Using a variety of
approaches, ranging from the algebraic techniques of BCFT to the
lagrangian description based on the boundary WZW model, it has been
possible to analyse in a detailed and systematic fashion what are the
consistent D-brane configurations in a given group manifold and how
they can be classified.

In this talk, based on \cite{SDnotes,Q0,FSrc,SU3,SD2notes}, we present
a geometric approach to the study of D-branes in group manifolds (that
is, in WZW models).  In Section 2 we describe how the classical
geometry of these D-branes can be determined directly from the gluing
conditions.  In Section 3 we discuss the boundary WZW model and how it
can be thought of as providing a lagrangian description of D-branes in
group manifolds.  In particular, we exhibit the two-form field defined
on the D-brane and the quantisation conditions obtained by requiring
that the path integral be well defined.  Finally, in Section 4, we
describe the classical and quantum moduli spaces of consistent D-brane
configurations in the Lie group $\SU_3$.  In particular, we show that
(twisted) D-brane configurations are in one-to-one correspondence with
the integrable highest weight (IHW) representations of the (twisted)
affine Lie algebra $\widehat\su(3)^{(2)}_k$.

\section{Symmetric D-branes in group manifolds}

The simplest and best understood class of D-brane configurations is
obtained \cite{AS,FFFS,SDnotes} as solutions of the familiar gluing
conditions on the chiral currents of the WZW model
\begin{equation}\label{eq:gc}
J(z) = R \bar J(\bar z) \qquad\text{at the boundary,}
\end{equation}
where $R$ is a metric preserving Lie algebra automorphism.  These
gluing conditions describe \emph{symmetric} D-branes, that is,
configurations which preserve the maximal amount of symmetry of the
bulk theory; that is, conformal invariance plus (some of) the current
algebra.

Let us assume, for simplicity, that $G$ is a compact, connected,
simply-connected Lie group.  A D-brane in $G$ wraps a submanifold $Q$
of $G$ on which there is defined a two-form field $\omega$.  The
D-submanifold $Q$ can be determined directly from the gluing
conditions \eqref{eq:gc}, by using the following geometric
interpretation \cite{SDnotes}.  The boundary conditions satisfied by
an open string whose end lies on $Q$ are given by
\begin{equation*}
\d g = \tilde R(g)\bar\d g~,
\end{equation*}
where $\tilde R(g) = \rho_g R \lambda_{g^{-1}}$ is the point dependent
matrix of boundary conditions.  The tangent space to the D-brane $T_g
Q$ and its perpendicular complement $T_g Q^\perp$ are spanned by the
eigenvectors of $\tilde R(g)$ corresponding to the Neumann and
Dirichlet conditions, respectively.  In particular, one can show that
the tagent space to the D-brane
\begin{equation*}
T_g Q = \Im (\id + \tilde R(g))~,
\end{equation*}
is nothing but the tangent space $T_g\eC_r(g)$ to a twisted conjugacy
class \cite{SDnotes,FFFS}
\begin{equation*}
\eC_r (h) = \left\{ r(g)h g^{-1} \mid g\in G\right\}~,
\end{equation*}
where $r:G\to G$ is the Lie group automorphism induced by $R$.  This
shows that D-branes described by \eqref{eq:gc} wrap twisted conjugacy
classes in the group manifold $G$.  In the special case $R=\id$, one
obtains \cite{AS} the standard conjugacy classes of $G$.

The twisted conjugacy class $\eC_r(h)$ of an element $h$ of $G$ is
defined as the orbit of $h$ in $G$ under the twisted adjoint action
$\Ad^r_g:g\mapsto r(g)hg^{-1}$, for any $g$ in $G$.  Since the
stabiliser of $h$ is given, in this case, by its twisted centraliser
$\eZ_r(g)=\{g\in G \mid r(g)h=hg\}$, the twisted conjugacy class
$\eC_r(h)$ can be described as the homogeneous space
\begin{equation}\label{eq:cc}
\eC_r(h)\cong G/{\eZ_r(g)}~.
\end{equation}

\section{The boundary WZW model}

The boundary WZW \cite{KlS,Gaw,FSrc} model can be thought of as a
lagrangian description for D-branes in group manifolds.  In this
framework, a D-brane is described by a submanifold $\iota:Q\to G$
together with a two-form $\omega$ on $Q$ such that $\iota^* H=
d\omega$, where $H = {1/6}~\la\theta,[\theta,\theta]\ra$ denotes the
three-form field on the target group manifold, $\theta$ is the
left-invariant Maurer-Cartan one-form on $G$, and $\la-,-\ra$ is an
invariant metric on the Lie algebra $\fg$ of $G$.  The classical
dynamics of an open string whose end lies on this D-brane is governed
by the action
\begin{equation}\label{eq:bwzw}
I = \int_\Sigma \left<g^{-1}\partial g,g^{-1}\Bar\partial g\right> +
  \int_M H - \int_D \omega~,
\end{equation}
where $M$ is a $3$-dimensional submanifold of $G$ with boundary
$\partial M = g(\Sigma) + D$, and $D$ is a $2$-dimensional submanifold
of $Q$.  There exists a homological obstruction \cite{FSrc} to the
existence of $M$ which is measured by the relative homology class of
$g(\Sigma)$ in $H_2(\G,Q)$.

The action of the boundary WZW model \eqref{eq:bwzw} is constructed as
a natural generalisation of the standard WZW action; in particular, it
is invariant under the infinite-dimensional symmetry group generated
by the transformations
\begin{equation}\label{eq:sym}
g(z,\bar z)\mapsto \Omega(z)g(z,\bar z)\bar\Omega(\bar z)^{-1}~.
\end{equation}
The parameters $\Omega(z)$ and $\bar\Omega(\bar z)$ of these
transformations satisfy $\bar\d\Omega = \d\bar\Omega = 0$ and are such
that \eqref{eq:sym} preserves the worldsheet boundary
$g(\Sigma)\subset\eC_r$; this latter property is encoded in the
condition $\Omega(z) = r\cdot\bar\Omega(\bar z)$, at the boundary
$\d\Sigma$.  In terms of the conserved chiral currents $J(z)$ and
$\bar J(\bar z)$, this gives rise to the gluing conditions
\eqref{eq:gc}.

The two-form field $\omega$ is uniquely determined by the symmetry
requirement \eqref{eq:sym}, being given \cite{Q0} (see also
\cite{AS,Gaw} for the case $R=\id$) by
\begin{equation}
  \label{eq:omega}
  \omega = -\half~\langle g^{-1}dg\ , \frac{\id + \Ad_{g^{-1}}R}{\id -
    \Ad_{g^{-1}}R}\ g^{-1}dg\rangle~.
\end{equation}
One can easily check that the right hand side is well defined on the
twisted conjugacy class $\eC_r$, and that \eqref{eq:omega} defines a
field which satisfies $d\omega = \iota^* H$.

In the case of the standard WZW model, we recall that the cancellation
of the global worldsheet anomaly imposes that the period of the
three-form field $H$ over any $3$-cycle in $H_3(\G)$ be quantised,
that is $[H]/2\pi\in H^3(\G,\Z)$.  Similarly, in the case of the
boundary WZW model, the condition that the path integral be single
valued (that is, that it not depend on the choice of $M$) imposes that
the global worldsheet anomaly vanish, which translates into
\begin{equation}\label{eq:qc}
  \int_N H - \int_{\d N} \omega \ \in 2\pi\Z~,
\end{equation}
for any relative $3$-cycle $(N,\d N)$ in $H_3(G,\eC)$.  In other
words, $[(H,\omega)]/2\pi$ must define a class in $H^3(\G,\eC_r;\Z)$.

\section{Example: Symmetric D-branes in $SU_3$}

Let us now apply the formalism described in the previous sections in
order to determine the consistent symmetric D-branes in $\SU_3$, for
which we have two distinct classes of solutions: untwisted branes,
characterised by $r=\id$, and twisted branes, defined by $r=\tau$,
where $\tau$ is the Dynkin diagram automorphism of $\SU_3$.

\subsection{Untwisted branes}

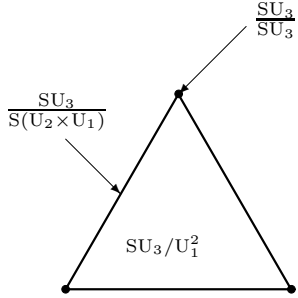
\begin{figure}[h!]
  \begin{center}
    \begin{picture}(45,40)
      \put(10,1){%
        \put(0,0){\circle*{1}}
        \put(30,0){\circle*{1}}
        \put(15,25.9808){\circle*{1}}
        \thicklines
        \path(0,0)(30,0)(15,25.9808)(0,0)
        \thinlines
        \put(8,5){$\scriptstyle \SU_3/\U_1^2$}
        \put(0,20){\vector(1,-1){7.35}}
        \put(-8,23){$\frac{\SU_3}{\mathrm{S}(\U_2\times \U_1)}$}
        \put(24,34.9808){\vector(-1,-1){9}}
        \put(25,35){$\frac{\SU_3}{\SU_3}$}
        }
    \end{picture}
    \caption{Moduli space of conjugacy classes of $\SU_3$}
    \label{fig:McSU3}
  \end{center}
\end{figure}

Conjugacy classes in a group $G$ are parametrised by the maximal torus
$T$ of $G$, modulo the Weyl group.  Standard group theory tells us
that the classical moduli space of D-branes in $\SU_3$, denoted here
by $\eM_{cl}(\SU_3,\id)$, which is the same as the space of conjugacy
classes of $\SU_3$, can be identified with the fundamental domain of
the extended Weyl group in the Cartan subalgebra $\ft$, which is given
by the (solid) equilateral triangle
\begin{equation}\label{eq:clmcc}
\eM_{cl}(\SU_3,\id) = \{ X\in\ft \mid 0 \leq \alpha_i(X) \leq 1~,\
                       i=1,2,3\}~,
\end{equation}
where $\alpha_1$, $\alpha_2$ and $\alpha_3=\alpha_1+\alpha_2$ are the
positive roots of $\SU_3$.  The interior points in $\eM_{cl}$ are
regular, and give rise to $6$-dimensional conjugacy classes of the
form $\SU_3/\U_1^2$.  If we now consider an element $X$ in $\ft$ which
belongs to one of the edges of $\eM_{cl}$, this describes an element
$h=\exp(X)$ in the maximal torus of $\SU_3$, whose centraliser
includes a $\SU_2^{\alpha_i}$ subgroup, for some $i$.  Thus the
boundary points belonging to the three edges are singular, giving rise
to $4$-dimensional conjugacy classes of the form
$\SU_3/\mathrm{S}(\U_2\times\U_1)$.  Finally, the three vertices
corresponding to the three central elements of $\SU_3$ describe
point-like D-branes of the form $\SU_3/\SU_3$.  We thus obtain the
space of classical symmetric D-branes represented in
Figure~\ref{fig:McSU3}.

By working out the quantisation conditions \eqref{eq:qc} one obtains
that the quantum moduli space $\eM_q(\SU_3,\id)$ of D-branes in $\SU_3$
at level $k$ is given by 
\begin{equation}\label{eq:qms1}
\eM_q(\SU_3,\id) = \left\{X\in\fh \mid k\alpha_i(X)\in\Z,\ 0
               \leq k\alpha_i(X)\leq k,\ i=1,2,3 \right\}~,
\end{equation}
which proves \cite{Gaw} that the set of consistent symmetric D-brane
configurations at level $k$ is in one-to-one correspondence with the
set of IHW representations of the corresponding affine Lie algebra
$\widehat\su(3)^{(1)}_k$.

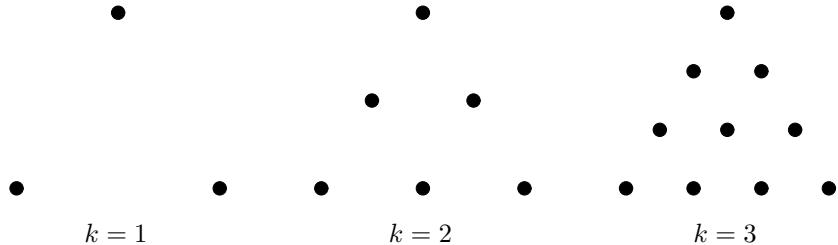
\begin{figure}[h!]
  \setlength{\unitlength}{0.9mm}
  \begin{center}
    \begin{picture}(130,40)
      \put(5,10){%
        \put(0,0){\circle*{2}}
        \put(30,0){\circle*{2}}
        \put(15,25.9808){\circle*{2}}
        \put(10,-8){\makebox{$k=1$}}
        }
      \put(50,10){%
        \put(0,0){\circle*{2}}
        \put(15, 0){\circle*{2}}
        \put(7.5, 12.9904){\circle*{2}}
        \put(22.5, 12.9904){\circle*{2}}
        \put(30,0){\circle*{2}}
        \put(15,25.9808){\circle*{2}}
        \put(10,-8){\makebox{$k=2$}}
        }
      \put(95,10){%
        \put(0,0){\circle*{2}}
        \put(10,0){\circle*{2}}
        \put(20,0){\circle*{2}}
        \put(30,0){\circle*{2}}
        \put(5, 8.66025){\circle*{2}}
        \put(10, 17.3205){\circle*{2}}
        \put(15, 8.66025){\circle*{2}}
        \put(20, 17.3205){\circle*{2}}
        \put(25, 8.66025){\circle*{2}}
        \put(15,25.9808){\circle*{2}}
        \put(10,-8){\makebox{$k=3$}}
        }
    \end{picture}
    \caption{Quantum moduli space for $\SU_3$ for lowest values of
    the level $k$.}
    \label{fig:MqSU3}    
  \end{center}
\end{figure}

The space of untwisted D-brane configurations in $\SU_3$ for the first
few values of the level $k$ is represented in Figure~\ref{fig:MqSU3}.
At a given level $k$ we have $3$ point-like, $3(k-1)$ $4$-dimensional
and $\half(k-1)(k-2)$ $6$-dimensional symmetric D-branes.  We also see
that the $4$-dimensional conjugacy classes are characterised by
quantum numbers $(\lambda_1,\lambda_2)$ with either one of the
$\lambda$'s being equal to zero or $\lambda_1+\lambda_2 = k$; the
point-like conjugacy classes are described by $(0,0)$, $(0,k)$,
$(k,0)$.  In particular, the lower-dimensional conjugacy classes
dominate the spectrum of D-branes until $k=9$.

\subsection{Twisted branes}

Let us now turn to the case of twisted branes.  Twisted conjugacy
classes are parametrised by the maximal torus $T^\tau$ of the fixed
point subgroup $\SU_3^\tau\cong\SO_3$, modulo the twisted Weyl group
$W_\tau$ of $\SU_3$.  In order to understand $W_\tau$ and determine
the space of twisted conjugacy classes, one needs to make an incursion
\cite{Siebenthal, Wendt} into the theory of non-connected Lie groups.
One obtains \cite{SU3} in this fashion a nice description of the
classical moduli space of twisted branes in $\SU_3$ in terms of the
classical moduli space of untwisted branes in $\SU_3^\tau$.  More
precisely, we have
\begin{equation}\label{eq:clmtcc}
\eM_{cl}(\SU_3,\tau) = \left\{ X\in\ft^\tau \mid 0\leq \bar\alpha(X)\leq
                        \tfrac{1}{4} \right\}~,
\end{equation}
where $\bar\alpha=\half(\alpha_1+\alpha_2)$ is the root of
$\SU_3^\tau$. 

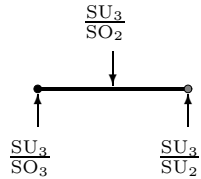
\begin{figure}[h!]
  \begin{center}
    \begin{picture}(40,30)
      \Thicklines
      \put(10.5,15){\line(1,0){19}}
      \thinlines
      \put(30,15){\shade\circle{1}}
      \put(10,15){\circle*{1}}
      \put(20,20){\vector(0,-1){4.5}}
      \put(16,23){$\frac{\SU_3}{\SO_2}$}
      \put(10,10){\vector(0,1){4.3}}
      \put(6,5){$\frac{\SU_3}{\SO_3}$}
      \put(30,10){\vector(0,1){4.3}}
      \put(26,5){$\frac{\SU_3}{\SU_2}$}
    \end{picture}
    \caption{Moduli space of twisted conjugacy classes of $\SU_3$}
    \label{fig:MtcSU3}
  \end{center}
\end{figure}

The resulting space of classical twisted D-branes in $\SU_3$ is
described in Figure~\ref{fig:MtcSU3}.  The point $\bar\alpha(X)=0$ is
singular in $\SO_3$ and $\tau$-singular (see \cite{SU3}) in $\SU_3$,
giving rise to a $5$-dimensional twisted D-brane of the form
$\SU_3/\SO_3$.  The other endpoint $4\bar\alpha(X)=1$ is regular in
$\SO_3$, but $\tau$-singular in $\SU_3$.  The corresponding twisted
class is also $5$-dimensional, but has the form $\SU_3/\SU_2$.
Finally, the interior points are $\tau$-regular and give rise to
$7$-dimensional twisted conjugacy classes of the form $\SU_3/\SO_2$.
Notice that in this case we have that the dimension of the twisted
conjugacy classes is always odd, due to the fact that the difference
between the ranks of $\SU_3$ and $\SO_3$ is odd.

\begin{figure}[h!]
  \setlength{\unitlength}{0.9mm}
  \begin{center}
    \begin{picture}(120,60)
      \put(5,50){%
        \put(0,0){\circle{2}}
        \put(20,0){\circle*{2}}
        \put(40,0){\circle{2}}
        \put(15,-8){\makebox{$k=1$}}
        }
      \put(75,50){%
        \put(0,0){\circle*{2}}
        \put(10,0){\circle{2}}
        \put(20,0){\circle*{2}}
        \put(30,0){\circle{2}}
        \put(40,0){\circle{2}}
        \put(15,-8){\makebox{$k=2$}}
        }
      \put(5,30){%
        \put(0,0){\circle{2}}
        \put(6.6667,0){\circle*{2}}
        \put(13.3333,0){\circle{2}}
        \put(20,0){\circle*{2}}
        \put(26.6667,0){\circle{2}}
        \put(33.3333,0){\circle{2}}
        \put(40,0){\circle{2}}
        \put(15,-8){\makebox{$k=3$}}
        }
      \put(75,30){%
        \put(0,0){\circle*{2}}
        \put(5,0){\circle{2}}
        \put(10,0){\circle*{2}}
        \put(15,0){\circle{2}}
        \put(20,0){\circle*{2}}
        \put(25,0){\circle{2}}
        \put(30,0){\circle{2}}
        \put(35,0){\circle{2}}
        \put(40,0){\circle{2}}
        \put(15,-8){\makebox{$k=4$}}
        }
      \put(5,10){%
        \put(0,0){\circle{2}}
        \put(4,0){\circle*{2}}
        \put(8,0){\circle{2}}
        \put(12,0){\circle*{2}}
        \put(16,0){\circle{2}}
        \put(20,0){\circle*{2}}
        \put(24,0){\circle{2}}
        \put(28,0){\circle{2}}
        \put(32,0){\circle{2}}
        \put(36,0){\circle{2}}
        \put(40,0){\circle{2}}
        \put(15,-8){\makebox{$k=5$}}
        }
      \put(75,10){%
        \put(0,0){\circle*{2}}
        \put(3.3333,0){\circle{2}}
        \put(6.6667,0){\circle*{2}}
        \put(10,0){\circle{2}}
        \put(13.3333,0){\circle*{2}}
        \put(16.6667,0){\circle{2}}
        \put(20,0){\circle*{2}}
        \put(13.3333,0){\circle{2}}
        \put(16.6667,0){\circle{2}}
        \put(23.3333,0){\circle{2}}
        \put(26.6667,0){\circle{2}}
        \put(30,0){\circle{2}}
        \put(33.3333,0){\circle{2}}
        \put(36.6667,0){\circle{2}}
        \put(40,0){\circle{2}}
        \put(15,-8){\makebox{$k=6$}}
        }
    \end{picture}
    \caption{Quantum moduli space for $\tau$-twisted D-branes in
      $\SU_3$ at level $k$ (black circles) compared with that of
      $\SU_3^\tau \protect\cong \SO_3$ (white circles) at level $4k$.}
    \label{fig:MqSU3SO3}
  \end{center}
\end{figure}
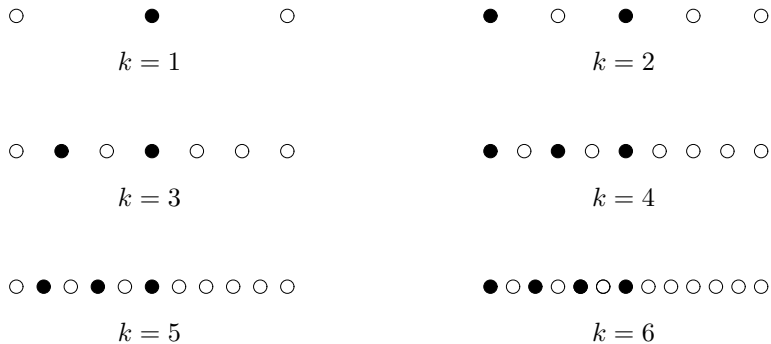

If we now impose the condition \eqref{eq:qc} for the cancellation of
the global worldsheet anomaly, we obtain that the quantum moduli space
of twisted D-branes in $\SU_3$ is given by
\begin{equation}\label{eq:qmstau}
\eM_q(\SU_3,\tau) = \begin{cases}
\{X\in\ft^\tau \mid 4k\bar\alpha(X) = 1,3,...,k\}~, &\text{for}\quad
                                      k\ \text{odd}~,\\ 
\{X\in\ft^\tau \mid 4k\bar\alpha(X) = 0,2,...,k\}~, &\text{for}\quad
                                      k\ \text{even}~.
                    \end{cases}
\end{equation}
The states corresponding to the first few values of the level are
represented in Figure~\ref{fig:MqSU3SO3}.  At a given odd level $k$ we
have $\half(k-1)$ $7$-dimensional and one $5$-dimensional branes,
whereas for $k$ even we have $(\half k - 1)$ $7$-dimensional and two
$5$-dimensional branes.

A careful comparison of the quantisation conditions \eqref{eq:qc} for
the twisted branes with the spectrum \cite{Kac} of IHW representations
of the twisted affine Lie algebra $\widehat\su(3)^{(2)}_k$ reveals
\cite{SU3} that the admissible twisted D-brane configurations in
$\SU_3$ are in one-to-one correspondence with the IHW representations
of the corresponding twisted affine Lie algebra
$\widehat\su(3)^{(2)}_k$.

%

\begin{thebibliography}{10}

\bibitem{SDnotes}
S.~Stanciu, ``D-branes in group manifolds,'' {\em J. High Energy Phys.} {\bf
  01} (2000) 025. \texttt{arXiv:hep-th/9909163}.

\bibitem{Q0}
S.~Stanciu, ``A note on {D}-branes in group manifolds: flux quantisation and
  {D}0-charge,'' {\em J. High Energy Phys.} {\bf 10} (2000) 015.
  \texttt{arXiv:hep-th/0006145}.

\bibitem{FSrc}
J.~Figueroa-O'Farrill and S.~Stanciu, ``D-brane charge, flux quantisation and
  relative (co)homology,'' {\em J. High Energy Phys.} {\bf 01} (2001) 006.
  \texttt{arXiv:hep-th/0008038}.
  \texttt{arXiv:hep-th/0006145}.

\bibitem{SU3}
S.~Stanciu, ``An illustrated guide to D-branes in $\SU_3$,''
\texttt{arXiv:hep-th/0111221}.

\bibitem{SD2notes}
S.~Stanciu, ``D-branes in group manifolds {II}.'' to appear.

\bibitem{AS}
A.~Alekseev and V.~Schomerus, ``D-branes in the {WZW} model,'' {\em Phys. Rev.}
  {\bf D60} (1999) 061901. \texttt{arXiv:hep-th/9812193}.

\bibitem{FFFS}
G.~Felder, J.~Fr\"ohlich, J.~Fuchs, and C.~Schweigert, ``The geometry of {WZW}
  branes,'' {\em J. Geom. Phys.} {\bf 34} (2000) 162--190.
  \texttt{arXiv:hep-th/9909030}.





\bibitem{KlS}
C.~Klim{\v c}ik and P.~Severa, ``Open strings and {D-branes} in {WZNW}
  models,'' {\em Nuc. Phys.} {\bf B488} (1997) 653--676.
  \texttt{arXiv:hep-th/9609112}.

\bibitem{Gaw}
K.~Gaw\c{e}dzki, ``Conformal field theory: a case study.''
  \texttt{arXiv:hep-th/9904145}.

\bibitem{FSNW}
J.~Figueroa-O'Farrill and S.~Stanciu, ``More {D}-branes in the {N}appi-{W}itten
  background,'' {\em J. High Energy Phys.} {\bf 01} (2000) 024.
  \texttt{arXiv:hep-th/9909164}.

\bibitem{Kac}
V.~Kac, {\em Infinite dimensional Lie algebras}.
\newblock Cambridge University Press, third~ed., 1990.


\bibitem{Siebenthal}
J.~de~Siebenthal, ``Sur les groupes de {L}ie compacts non connexes,'' {\em
  Comment. Math. Helv.} {\bf 31} (1956) 41--89.

\bibitem{Wendt}
R.~Wendt, ``Weyl's character formula for non-connected {L}ie groups and orbital
  theory for twisted affine {L}ie algebras.'' \texttt{arXiv:math.RT/9909059}.


\end{thebibliography}
%
\providecommand{\href}[2]{#2}\begingroup\raggedright\endgroup

\end{document}